\documentclass[pre,twocolumn,superscriptaddress,showpacs,floatfix]{revtex4}
\usepackage{graphicx}
\usepackage{amssymb}
\usepackage{amsmath}
\usepackage{amstext}
\usepackage{latexsym}
\usepackage[pdfmark, colorlinks,  citecolor=red, linkcolor=blue,urlcolor=blue]{hyperref}

\newcommand{\beq}{\begin{equation}}
\newcommand{\eeq}{\end{equation}}
\newcommand{\barr}{\begin{eqnarray}}
\newcommand{\earr}{\end{eqnarray}}    
\newcommand{\ket}[1]{\left\vert#1\right\rangle}
\newcommand{\bra}[1]{\left\langle#1\right\vert}

\newcommand{\abs}[1]{\left\vert #1 \right\vert}

\begin{document}
\title{Conservative chaotic map as a model of quantum many-body environment}

\author{Davide Rossini}
\email{d.rossini@sns.it}
\homepage{http://qti.sns.it}
\affiliation{NEST-CNR-INFM \& Scuola Normale Superiore, Piazza dei Cavalieri 7, I-56126 Pisa, Italy} 
\author{Giuliano Benenti}
\email{giuliano.benenti@uninsubria.it}
\homepage{http://scienze-como.uninsubria.it/complexcomo/}
\affiliation{Center for Nonlinear and Complex Systems,
Universit\`a degli Studi dell'Insubria, Via Valleggio 11,
22100 Como, Italy} \affiliation{Istituto
Nazionale di Fisica Nucleare, Sezione di Milano and CNISM}
\author{Giulio Casati}
\email{giulio.casati@uninsubria.it}
\affiliation{Center for Nonlinear and Complex Systems,
Universit\`a degli Studi dell'Insubria, Via Valleggio 11,
22100 Como, Italy} \affiliation{Istituto
Nazionale di Fisica Nucleare, Sezione di Milano and CNISM}
\affiliation{Department of Physics, National University of
Singapore, Singapore 117542, Republic of Singapore}

\date{\today}
\begin{abstract}
  We study the dynamics of the entanglement between two qubits 
  coupled to a common chaotic environment, described 
  by the quantum kicked rotator model.
  We show that the kicked rotator, which is a single-particle
  deterministic dynamical system, can reproduce
  the effects of a pure dephasing many-body bath.
  Indeed, in the semiclassical limit the interaction with the
  kicked rotator can be described as a random phase-kick, so that
  decoherence is induced in the two-qubit system.
  We also show that our model can efficiently simulate 
  non-Markovian environments.
\end{abstract}


\maketitle

\section{Introduction} \label{sec:introduction}

Real physical systems are never isolated and the coupling of the
system to the environment leads to decoherence. This process can 
be understood as the loss of quantum information, initially
present in the state of the system, when non-classical
correlations (entanglement) establish between the system and
the environment. On the other
hand, when tracing over the environmental degrees of freedom,
we expect that the entanglement between internal degrees
of freedom of the system is reduced or even destroyed.
Decoherence theory has a fundamental interest, 
since it provides explanations of the emergence of classicality
in a world governed by the laws of quantum mechanics \cite{zurek}.
Moreover, it is a threat to the
actual implementation of any quantum computation and
communication protocol \cite{chuang,qcbook}. 
Indeed, decoherence invalidates the
quantum superposition principle, which is at the heart of the
potential power of any quantum algorithm. A deeper understanding
of the decoherence phenomenon seems to be essential to develop
quantum computation technologies. 

The environment is usually described as a many-body quantum system.
The best-known model is the Caldeira-Leggett model \cite{caldeiraleggett},
in which the environment is a bosonic bath consisting of infinitely
many harmonic oscillators at thermal equilibrium. 
More recently, first studies of the role played by chaotic 
dynamics in the decoherence process have been carried out
\cite{park,blume03,lee,saraceno,seligman}.
These investigations mainly focused on the differences between the 
processes of decoherence induced by chaotic and regular environments.

In this paper, we address the following question: could
the many-body environment be substituted, without changing the 
effects on system's dynamics, by a closed
deterministic system with a small number of degrees of freedom
yet chaotic? In other words, can the complexity of the environment
arise not from being many-body but from chaotic dynamics?
In this paper, we give a positive answer to this question.
We consider two qubits coupled to a 
\emph{single particle, fully deterministic, conservative 
chaotic ``environment''}, described by the kicked rotator model. 
We show that, due to this system-environment interaction, 
the entropy of the system increases. At the same time, the
entanglement between the two qubits decays, thus illustrating
the loss of quantum coherence. We show that the evolution in
time of the two-qubit entanglement is in good agreement with the 
evolution obtained in a pure dephasing stochastic model.
Since this pure dephasing decoherence mechanism can be derived
in the framework of the Caldeira-Leggett model \cite{palma},  
we have established a direct link between the effects
of a many-body environment and of a chaotic single-particle 
environment.

Finally, we show that also non-Markovian effects naturally 
appear in our model. This is interesting since such effects
are relevant in several implementations of quantum information
protocols \cite{esteve}, for instance with solid-state devices due to the 
coupling of the qubits to ``slow'' modes such as phonons.
Moreover, non-Markovian effects, apart from some exceptions \cite{daffer}, 
are usually very hard for analytic treatment. On the other hand, 
they can be included in models similar to the one discussed in the 
present paper with no additional computational cost with respect to
the Markovian case.   

The structure of the paper is as follows: in Sec.~\ref{sec:model}
our model of a chaotic environment is introduced; in Sec.~\ref{sec:decoherence}
both the entropy production and the entanglement decay are investigated, and
the results compared with those obtained in the phase-kick model; 
in Sec.~\ref{sec:memory} non-Markovian effects are analyzed; 
our conclusions are drawn in Sec.~\ref{sec:conclusions}.
The Bloch representation of the phase-kick map and the short time
decay of the relevant bath (kicked rotator) correlation functions are
discussed in Appendix \ref{app1} and Appendix \ref{app2}, respectively.

\section{The model} \label{sec:model}

We consider two qubits coupled to a
quantum kicked rotator. The overall system is 
governed by the Hamiltonian
\beq
\hat{H} = \hat{H}^{(1)} + \hat{H}^{(2)} +
\hat{H}^{(\mathrm{kr})} + \hat{H}^{(\mathrm{int})} \, ,
\label{eq:hammodel} \eeq
where $\hat{H}^{(i)} = \omega_i \, \hat{\sigma}_x^{(i)}$ ($i=1,2$)
describes the free evolution of the two qubits,
\beq
\hat{H}^{(\mathrm{kr})} = \frac{\hat{n}^2}{2} + k \cos(\hat{\theta})
\sum_j \delta(\tau-jT)
\label{eq:krotator}
\eeq
the quantum kicked rotator, and
\beq
\hat{H}^{(\mathrm{int})} = \epsilon \: ( \hat{\sigma}_z^{(1)} +
\hat{\sigma}_z^{(2)}) \cos(\hat{\theta}) \sum_j \delta(\tau-jT)
\eeq
the interaction between the qubits and the kicked rotator;
$\hat{\sigma}_\alpha^{(i)}$ ($\alpha = x,y,z$) denote the Pauli matrices
of the $i$-th qubit.
Both the cosine potential in $\hat{H}^{(\mathrm{kr})}$ and 
the interaction $\hat{H}^{(\mathrm{int})}$ are switched on and off 
instantaneously (kicks) at regular time intervals $T$.
We consider the two qubits as an open quantum system and the kicked 
rotator as their \emph{common} environment.
Note that we chose noninteracting qubits as we want their entanglement
to be affected exclusively by the coupling to the environment.

The evolution from time $tT$ (prior to the $t$-th kick) to time $(t+1)T$ 
(prior to the $(t+1)$-th kick) of the kicked rotator in the classical limit
is described by the Chirikov standard map:
\beq
\left\{ \begin{array}{l}
\bar{n} = n + k \sin \theta, \\
\bar{\theta} = \theta + T \bar{n},
\hspace{1.cm} (\textrm{mod } 2 \pi),
\end{array} \right. \label{eq:chirikov}
\eeq
where $(n,\theta)$ are conjugated momentum-angle variables.
Hereafter $t=\tau/T$ denotes the discrete time, measured in number of kicks.
By rescaling $n \to p = T n$, the dynamics of Eqs.~\eqref{eq:chirikov}
is seen to depend only on the parameter $K = kT$.
For $K=0$ the motion is integrable; when $K$ increases,
a transition to chaos of the KAM type is observed.
The last invariant KAM torus is broken for $K \approx  0.97$.
If $K \sim 1$ the phase space is mixed (simultaneous presence of
integrable and chaotic components).
If $K$ increases further, the stability islands progressively reduce
their size; for $K \gg 1$ they are not visible any more.
In this paper, we always consider map (\ref{eq:chirikov}) on the torus
$0 \leq \theta < 2 \pi$, $- \pi \leq p < \pi$. 
In this case, the Chirikov standard map describes the stroboscopic dynamics of 
a \emph{conservative} dynamical system with two degrees of freedom which,
in the fully chaotic regime $K\gg 1$, relaxes, apart from quantum fluctuations, 
to the uniform distribution on the torus.

The Hilbert space of the global system is given by 
\beq
\mathcal{H} = \mathcal{H}^{(1)} \otimes \mathcal{H}^{(2)} \otimes
\mathcal{H}^{(kr)} \, ,
\eeq
where $\mathcal{H}^{(1)}$ and $\mathcal{H}^{(2)}$ are the two-dimensional
Hilbert spaces associated to the two qubits, and
$\mathcal{H}^{(kr)}$ is the $N$-dimensional space for the kicked rotator.

The time evolution generated by Hamiltonian 
(\ref{eq:hammodel}) in one kick is described by the operator
\beq
\begin{array}{rl}
\hat{U} = & \exp \big[  -i \big( k + \epsilon (\hat{\sigma}_z^{(1)} +
\hat{\sigma}_z^{(2)}) \big) \cos(\hat{\theta}) \big]  \\
& \times\, \exp \big[ -i T \frac{\hat{n}^2}{2} \big]
\exp \big( -i \, \delta_1 \, \hat{\sigma}_x^{(1)} \big) 
\exp \big( -i \, \delta_2 \, \hat{\sigma}_x^{(2)} \big).
\end{array} \label{eq:kickedevol}
\eeq
The effective Planck constant is $\hbar_{\textrm{eff}}=T = 2 \pi /N$,
where $N$ is the number of quantum levels used to describe the kicked rotator;
$\delta_1 = \omega_1 T , \: \delta_2 = \omega_2 T$;
$\epsilon$ is the coupling strength between the qubits and the environment.
The classical limit $\hbar_{\textrm{eff}} \to 0$ is obtained by taking
$T \to 0$ and $k \to \infty$, in such a way that $K = kT$ is kept fixed.

\section{Loss of coherence induced by a chaotic environment}
\label{sec:decoherence}

We are interested in the case in which the environment (the kicked rotator) is chaotic
(we consider the kicked rotator with $K \gg 1$).
The two qubits are initially prepared in a maximally entangled state,
such that they are disentangled from the environment.
Namely, we suppose that at $t=0$ the system is in the state
\beq
\ket{\Psi_0} = \ket{\phi^+} \otimes \ket{\psi_0} \, ,
\label{eq:initial}
\eeq
where $\ket{\phi^+} = \frac{1}{\sqrt{2}} \left( \ket{00} + \ket{11} \right)$
is a Bell state (the particular choice of the initial maximally entangled
state is not crucial for our purposes), and
$\ket{\psi_0} = \sum_n c_n \ket{n}$ is a generic state of
the kicked rotator, with $c_n$ random coefficients such that
$\sum_n \abs{c_n}^2 =1$, and $\ket{n}$ eigenstates of the momentum
operator.
The evolution in time of the global system (kicked rotator plus qubits)
is described by the unitary operator $\hat{U}$ defined in Eq.~\eqref{eq:kickedevol}.
Therefore, any initial pure
state $\ket{\Psi_0}$ evolves into another pure state
$\ket{\Psi (t)} = \hat{U}^t \ket{\Psi_0}$.
The reduced density matrix $\rho_{12} (t)$ describing the two qubits
at time $t$ is then obtained after tracing $\ket{\Psi (t)} \bra{\Psi (t)}$
over the kicked rotator's degrees of freedom.

In the following we will focus our attention on the time evolution
of the entanglement $E_{12}$ between the two qubits
and between them and the kicked rotator, measured by the reduced von Neumann
entropy $S_{12}$.
Clearly, for states like the one in Eq.~\eqref{eq:initial},
we have $E_{12} (0) = 1$, $S_{12} (0) = 0$.
As the total system evolves, we expect that $E_{12}$ decreases, while
$S_{12}$ grows up, thus meaning that the two-qubit system is
progressively losing coherence.

The entanglement of formation $E_{12}$ of a generic two-qubit state
$\rho_{12}$ can be evaluated following Ref.~\cite{wootters98}. 
First of all we compute the concurrence, defined as 
$C=\max ( \lambda_{1} - \lambda_{2} - 
\lambda_{3} - \lambda_{4} , 0 )$,
where the $\lambda_{i}$'s are the square roots of the eigenvalues 
of the matrix $R=\rho_{12} \tilde{\rho}_{12}$, in decreasing order.
Here $\tilde{\rho}_{12}$ is the spin flipped matrix of $\rho_{12}$, 
and it is defined by $\tilde{\rho}_{12}= 
(\sigma_{y} \otimes \sigma_{y}) \, \rho_{12}^{\star} \, 
(\sigma_{y} \otimes \sigma_{y})$ 
(note that the complex conjugate is taken in the computational
basis $\{ \vert 00 \rangle, \vert 01 \rangle, \vert 10 \rangle, 
\vert 11 \rangle \}$).
Once the concurrence has been computed, the entanglement 
of formation is obtained
as $E= h((1+\sqrt{1-C^{2}})/2)$, 
where $h$ is the binary
entropy function: $h(x)=-x\log_{2}x-(1-x)\log_{2}(1-x)$,
with $x=(1+\sqrt{1-C^{2}})/2$. 
The other quantity studied in our investigations is the
entanglement between the two qubits and
the kicked rotator, measured by the von Neumann entropy 
$S=-\mathrm{Tr} \, [ \rho_{12}\log_{2} \rho_{12}]$
of the reduced density matrix $\rho_{12}$.

If the kicked rotator is in the chaotic regime and in the semiclassical 
region $\hbar_{\textrm{eff}}\ll 1$, it is possible to drastically
simplify the description of the system in Eq.~\eqref{eq:hammodel}
by using the \emph{random phase-kick} approximation, in the framework
of the Kraus representation formalism.
Since, to a first approximation, the phases between two consecutive kicks
in the chaotic regime can be considered as uncorrelated, the interaction
with the environment can be simply modeled as a phase-kick rotating
both qubits through \emph{the same} random angle about the $z$-axis
of the Bloch sphere.
This rotation is described by the unitary matrix
\beq \label{eq:rotation}
\hat{R} (\theta) = \left[
\begin{array}{cc} e^{- i \epsilon \cos \theta} & 0 \\
0 & e^{i \epsilon \cos \theta} \end{array} \right] \otimes
\left[ \begin{array}{cc} e^{- i \epsilon \cos \theta} & 0 \\
0 & e^{i \epsilon \cos \theta} \end{array} \right],
\eeq
where the angle $\theta$ is drawn from a uniform random
distribution in $[ 0, 2\pi )$.
The one-kick evolution of the reduced density matrix $\rho_{12}$ is then
obtained after averaging over $\theta$:
\beq
\begin{array}{c}
\bar{\rho}_{12} = \frac{1}{2\pi} \int_{0}^{2 \pi} \hspace{-2mm} d \theta \,
\hat{R} (\theta) \, e^{-i \delta_2 \hat{\sigma}_x^{(2)}}
e^{-i \delta_1 \hat{\sigma}_x^{(1)}} 
\vspace{1mm} \\
\times\,\rho_{12} \,
e^{i \delta_1 \hat{\sigma}_x^{(1)}} e^{i \delta_2 \hat{\sigma}_x^{(2)}}
\hat{R}^{\dagger} (\theta). \label{eq:randomphase}
\end{array}
\eeq

\begin{figure}
\includegraphics[width=8.cm]{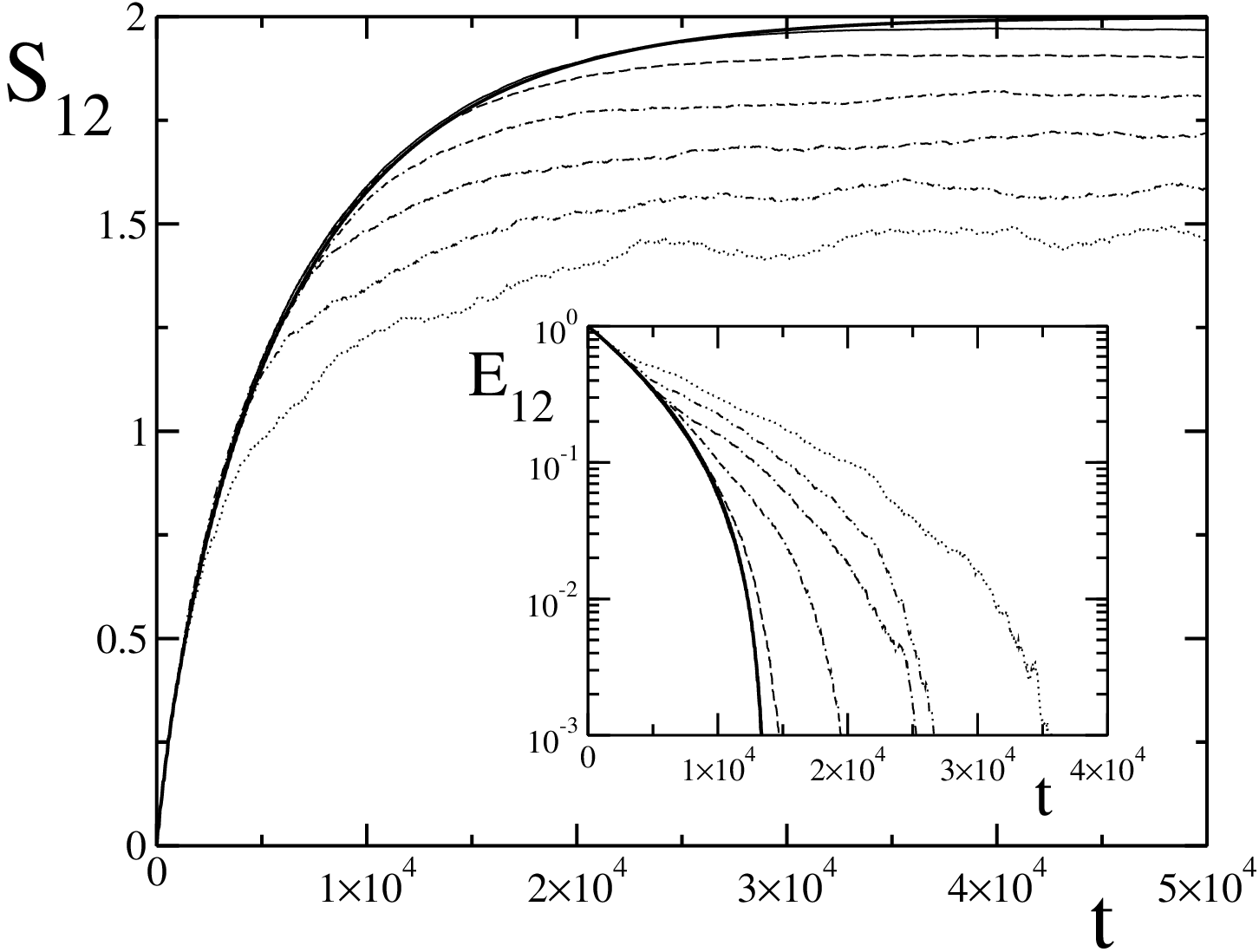}
\caption{Reduced von Neumann entropy $S_{12}$ (main figure) and 
entanglement $E_{12}$ (inset) as a function of time at $K \approx 99.73$,
$\delta_1=10^{-2}$, $\delta_2=\sqrt{2}\delta_1$, 
$\epsilon=8\times 10^{-3}$. The thin curves correspond
to different number of levels for the environment (the kicked rotator)
($N=2^9,2^{10},2^{11},2^{12},2^{13},2^{14}$ from bottom to
top in the main figure and vice versa in the inset).
The thick curves give the numerical results from the random phase model
(\ref{eq:randomphase}).}
\label{fig:confr_ES}
\end{figure}

In order to assess the validity of the random phase-kick approximation, 
we numerically investigate model (\ref{eq:hammodel}) in the 
classically chaotic regime $K\gg 1$ and in the 
region $\hbar_{\textrm{eff}}\ll 1$ in which the environment is a semiclassical 
object. Under these conditions, we expect that the 
time evolution of the entanglement can be accurately predicted
by the random phase model. Such expectation is confirmed by the numerical
data shown in  Fig.~\ref{fig:confr_ES}.
Even though differences between the
two models remain at long times due to the finite number $N$ of 
levels in the kicked rotator, such differences appear at later
and later times when $N\to \infty$ ($\hbar_{\textrm{eff}}\to 0$).
The parameter $K$ has been chosen much greater than one,
so that the classical phase space of the
kicked rotator can be considered as completely chaotic.
The actual value $K \approx 99.72676$ approximately corresponds to a zero
of the Bessel function $J_2 (K)$. This is to completely wipe off
memory effects between consecutive and next-consecutive kicks
(see Sec.~\ref{sec:memory}, Eqs.~\eqref{eq:corr1}-\eqref{eq:corr2}
for details).

We point out that the random phase model can be derived from the Caldeira-Leggett
model with a pure dephasing coupling 
$\propto ( \hat{\sigma}_z^{(1)} + \hat{\sigma}_z^{(2)}) \sum_k g_k \hat{q}_k$,
with $g_k$ coupling constant to the $k$-th oscillator of the environment, whose coordinate 
operator is $\hat{q}_k$ \cite{palma,braun}. This
establishes a direct link between our chaotic single-particle environment
and a standard many-body environment.

\begin{figure}
\includegraphics[width=8.cm]{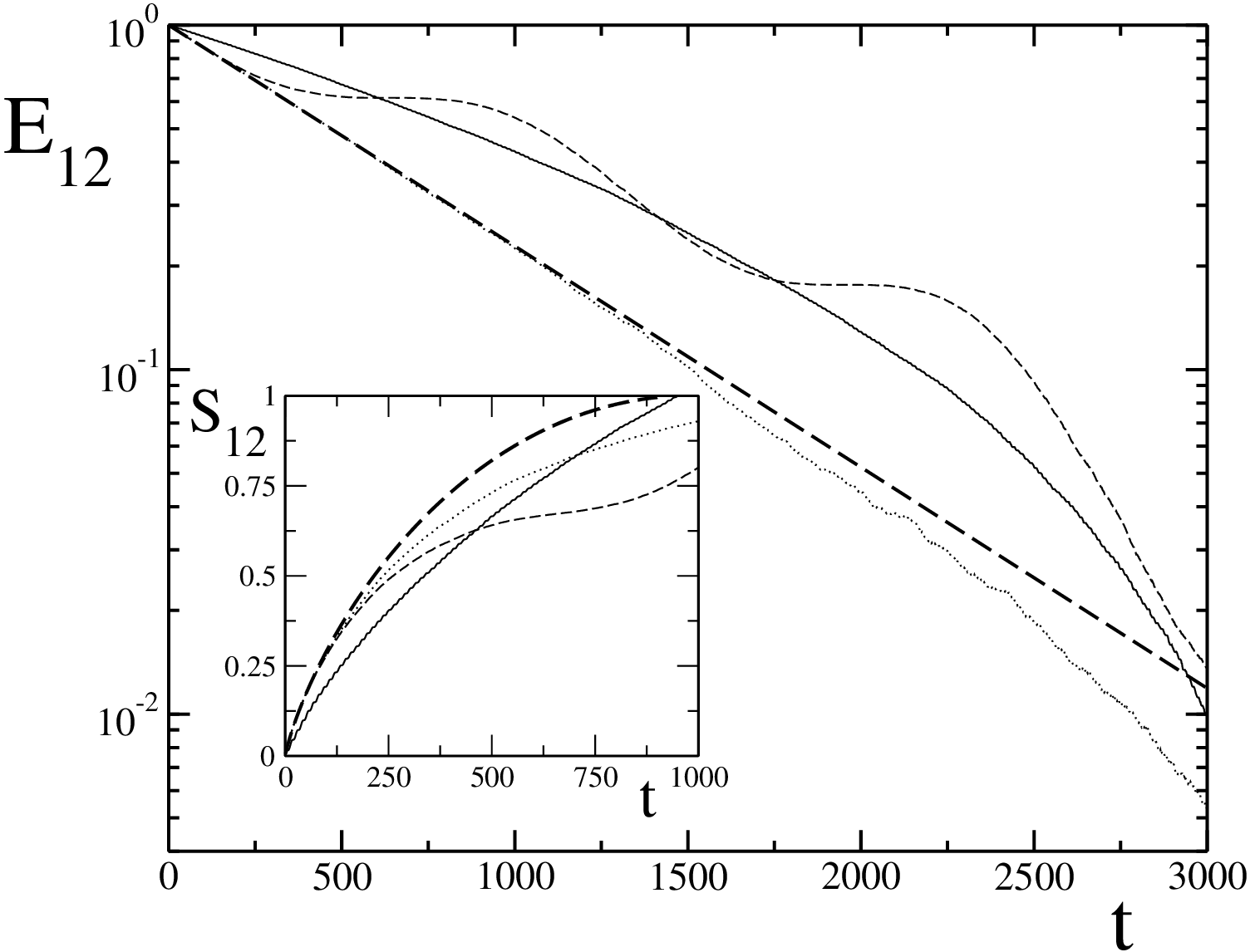}
\caption{Entanglement $E_{12}$ as a function of time for two qubits coupled to
a chaotic kicked rotator ($K \approx 99.73$) with $N=2^{14}$ levels
and coupling strength $\epsilon=0.016$.
The various curves are for different oscillation frequencies of
the qubits: $\delta_1 = \delta_2 / \sqrt{2} = 10^{-1}, 
\textrm{(solid curve)}, \: 10^{-3} \, \textrm{(dashed curve)}, \: 10^{-4}
\textrm{(dotted curve)}$. The thick dashed line gives the analytic estimate
\eqref{eq:Eestim}, valid in the regime $\delta_1,\delta_2 \ll \epsilon\ll 1$.
Inset: same as in the main figure but for 
the Von Neumann entropy $S_{12}$.
The thick dashed curve is the estimate \eqref{eq:Sestim}.}
\label{fig:ESdecay}
\end{figure}

\begin{figure}
\includegraphics[width=8.cm]{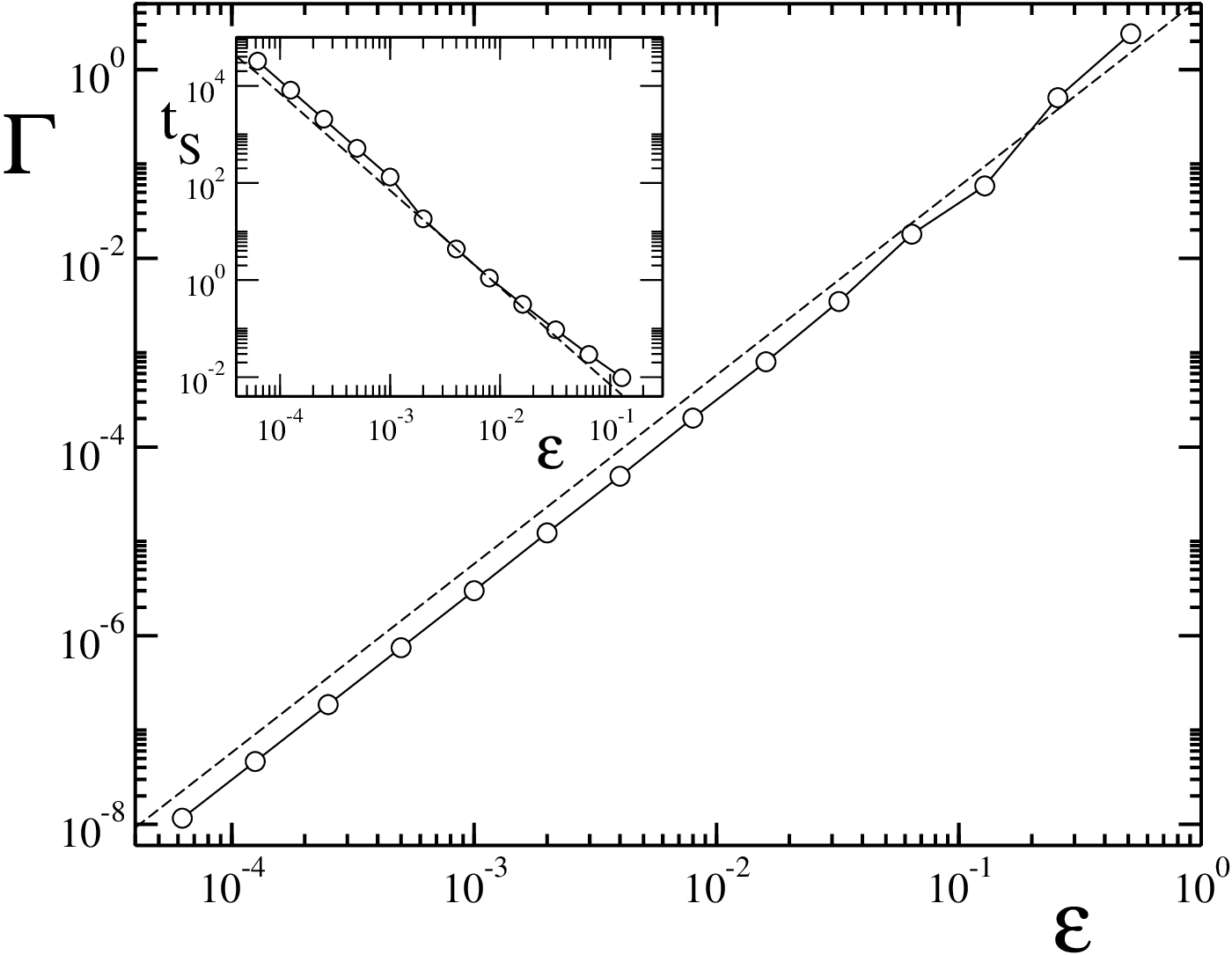}
\caption{Decay rates $\Gamma$ of the entanglement $E_{12}$ (main figure) 
and time scales $t_S$ for the Von Neumann entropy 
(obtained from $S_{12} (t_S) =0.002$) (inset)
as a function
of the coupling strength $\epsilon$ for $N=2^{14}$, $K \approx 99.73$,
$\delta_1= \delta_2 / \sqrt{2} = 10^{-2}$.
Both numerical data (circles) and  
the analytic predictions (dashed curves) given by Eq.~\eqref{eq:Eestim}
(main figure) and 
Eq.~\eqref{eq:Sestim} (inset) are shown.}
\label{fig:Grates}
\end{figure}

Numerical simulations show that, when the kicked rotator is chaotic,
the pairwise entanglement $E_{12}$ between the two qubits
decays exponentially at short times,
with a rate $\Gamma$ proportional to $\epsilon^2$:
\beq
E_{12} (t) \sim e^{-\alpha \epsilon^2 t} \, .
\label{eq:entdecay}
\eeq
This is clearly seen from Figs.~\ref{fig:ESdecay} and 
\ref{fig:Grates} \cite{lewenstein}.
In Fig.~\ref{fig:ESdecay}
we plot the decay of entanglement $E_{12}$ in time, for different 
values of the oscillation frequencies $\delta_1$ and $\delta_2$ of 
the two qubits, at a fixed value
of the coupling strength $\epsilon$ with the environment.
The short time decay of $E_{12}$ is always exponential
and superimposed to oscillations whose frequency is determined
by the internal dynamics of the two qubits~\cite{note_frequencies}.
Such oscillations can be clearly seen for 
$\delta_1=\delta_2/\sqrt{2}=10^{-3}$ (dashed curve),  
while their amplitude and period are too small to be seen 
on the scale of the figure for 
$\delta_1=\delta_2/\sqrt{2}=10^{-1}$ (solid curve).
Finally, oscillations are absent at  
$\delta_1=\delta_2/\sqrt{2}=10^{-4}$ (dotted curve), 
as in this case their 
period is longer than the time scale for entanglement decay.
Analogous remarks can be done about the oscillations in the 
entanglement production, measured by the reduced von Neumann entropy plotted
in the inset of Fig.~\ref{fig:ESdecay}.

An analytic estimate of the entanglement decay rate and of the 
entropy production can be derived
from the random phase-kick model, in the limiting case 
$\delta_1 , \delta_2 \ll \epsilon\ll 1$, that is, when the internal
dynamics of the two qubits can be neglected on the time scale
for entanglement decay.
Starting from Eq.~\eqref{eq:randomphase}, it is possible to explicitly
write down the map $\rho_{12} \to \bar{\rho}_{12}$ in the Bloch
representation, 
see Eqs.~\eqref{eq:mapequations} in App.~\ref{app1}.
Though in general these equations have to be solved numerically,
in App.~\ref{app1} we provide an explicit analytic solution
valid for $\delta_1 , \delta_2 \ll \epsilon\ll 1$.
In that case, for $\epsilon^2 t \ll 1$ we obtain:
\barr
E_{12} (t) & \simeq & 1 - \frac{4 \epsilon^2 t}{\ln 2}, \label{eq:Eestim}\\
S_{12} (t) & \simeq & \frac{2 \epsilon^2 t}{\ln 2}
\left( 1 - \ln (2 \epsilon^2 t) \right),
\label{eq:Sestim}
\earr
consistently with the numerically obtained behavior of
Eq.~\eqref{eq:entdecay} for small times, with $\alpha \approx 4 / \ln 2$.
The analytic estimates (\ref{eq:Eestim}) for $E_{12}$ and 
(\ref{eq:Sestim}) for $S_{12}$ are shown as thick dashed curves in 
Fig.~\ref{fig:ESdecay}. A very good agreement with
numerical data can be clearly seen for $\delta_1, \delta_2 \ll \epsilon\ll 1$.

In Fig.~\ref{fig:Grates} the decay rate $\Gamma$ of entanglement as a function
of $\epsilon$ is shown; circles represent numerical data,
while the dashed line is the analytic estimate of Eq.~\eqref{eq:Eestim}.
A dependence $\Gamma \propto \epsilon^2$ is found in both cases.
In the inset we plot the characteristic time scale $t_S$ for the
von Neumann entropy production. The time scale $t_S$ has been determined
from the condition $S_{12} (t_S) = C=0.002$ (the value chosen for $C$ is not crucial).
As before, circles stand for numerical data, while the dashed line
is obtained from Eq.~\eqref{eq:Sestim}.
From this figure we can conclude that the dependences of both 
$\Gamma$ and $t_S$ on 
the system-environment coupling strength $\epsilon$ are correctly 
reproduced by Eqs.~(\ref{eq:Eestim}) and (\ref{eq:Sestim}), even 
though the oscillations due to the internal system's dynamics cannot 
be reproduced by these analytic estimates. Finally, we point out that,
as clear from Fig.~\ref{fig:confr_ES}, in the semiclassical chaotic regime 
($\hbar_{\textrm{eff}}\ll 1$ and $K\gg 1$) both 
$\Gamma$ and $t_S$ can be reproduced with great accuracy by the random
phase-kick model. 

We would like to stress that the 
results discussed in this section do not depend on the initial
condition $|\psi_0\rangle$ in (\ref{eq:initial}),
provided that the kicked rotator is in
the chaotic regime. On the other hand, we have found that both the 
entanglement decay and the entropy production strongly depend on 
$|\psi_0\rangle$ in the integrable region $K<1$. This implies that
only in the chaotic regime a single particle can behave as
a dephasing environment.

Finally we notice that, contrary to other bath models~\cite{braun},
the random phase-kick model does not generate entanglement
between the two qubits.
This has been numerically checked for model (\ref{eq:hammodel})
both for initial pure separable states
and for separable mixtures.
Moreover we also checked that, starting from a generic two qubit
entangled state, the interaction with a chaotic memoryless environment
cannot increase entanglement. Namely, we considered $10^7$ random
initial conditions with $E_{12}(0) \neq 0$ and we found that,
already after $t \sim 100$ kicks, entanglement has been always lowered:
$E_{12}(t) < E_{12}(0)$.

\section{Non-Markovian effects} \label{sec:memory}

The model governed by Eq.~\eqref{eq:hammodel} is very convenient
for numerical investigations of decoherence.
It can also be used to study physical situations which
cannot be easily treated by means of analytic techniques
using many-body environments like in the Caldeira-Leggett model. 
For instance, one could simulate, with the same computational cost, 
more complex couplings (e.g. when the direction of the coupling 
changes in time) or non-Markovian environments.

\begin{figure}
\includegraphics[width=8.cm]{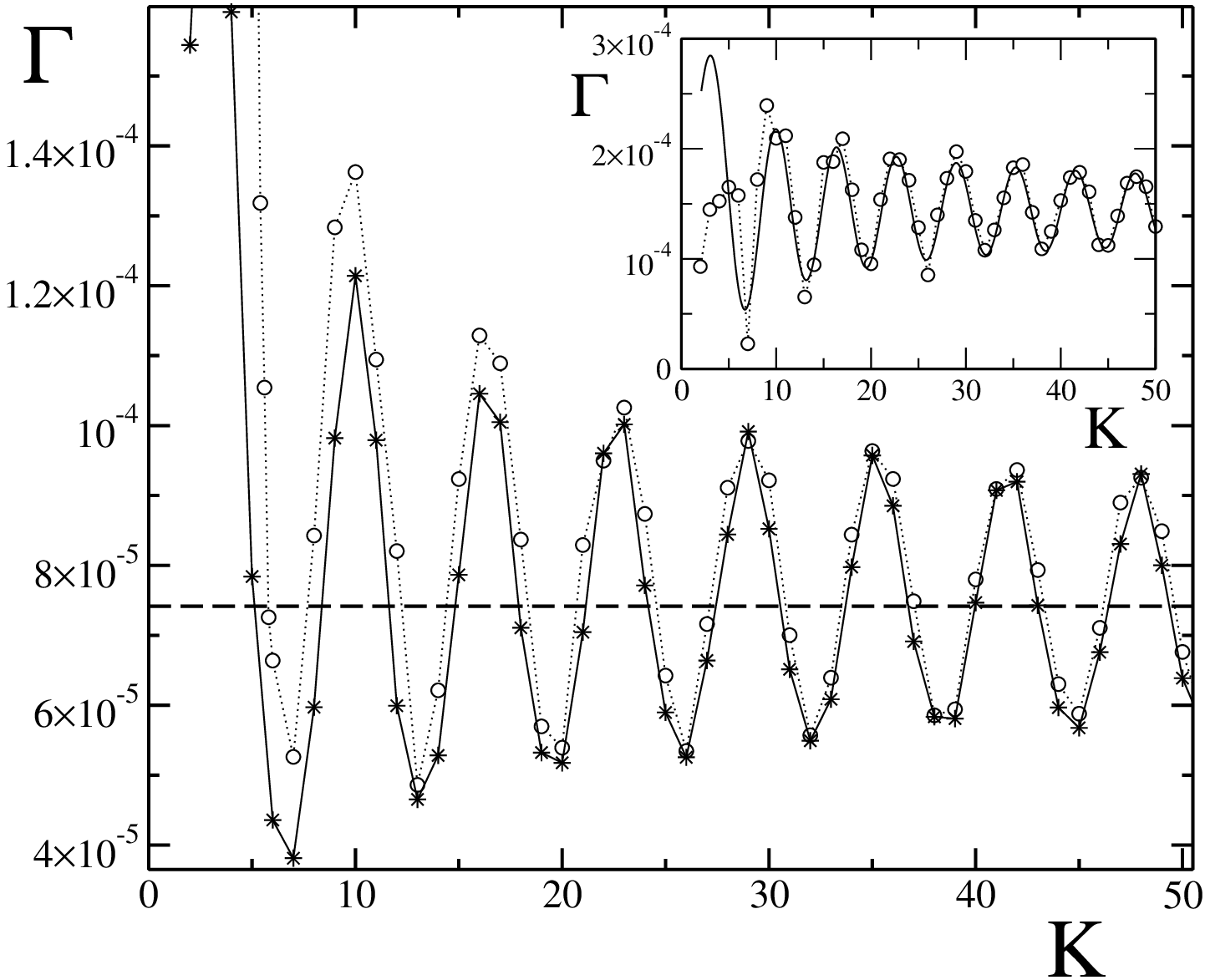}
\caption{Oscillations of the entanglement decay rate $\Gamma$ as a function of the 
classical chaos parameter $K$, at $\epsilon=5 \times 10^{-3},
\delta_1 = \delta_2 / \sqrt{2} = 10^{-2}$, $N=2^{14}$ (circles).
Stars show the results obtained from a quantum trajectory
approach (with average over 5000 trajectories) applied to the phase-kick model 
(\ref{eq:corrphase}).
Inset: entanglement decay rates as a function of $K$ for
$\epsilon=5 \times 10^{-3}$, $\delta_1 = \delta_2 = 0$, $N=2^{14}$ (circles).
The solid curve shows the analytic estimate \eqref{eq:corrrate}.}
\label{fig:krate}
\end{figure}

In this section, we show that memory effects naturally appear in our model,
outside the range of validity of the random phase approximation.
First of all, we point out that, given the coupling 
$\hat{H}^{(\mathrm{int})} \propto \cos \hat{\theta}$,
the bath correlation function relevant for the study of memory effects is 
$\langle \cos [\theta(t)] \cos [\theta(t^\prime)] \rangle$.
Even in the chaotic regime, the phases are not completely
uncorrelated: in the classical Chirikov standard map 
(\ref{eq:chirikov}) correlations
between cosines of the phases in two consecutive
kicks are zero, but the same correlations between phases of two
next-consecutive kicks do not vanish. As shown in Appendix \ref{app2}, we have
\barr
\label{eq:corr1}  \langle \cos \theta \cos \bar{\theta} \rangle & = & 0,
\\
\label{eq:corr2}  \langle \cos \theta \cos \bar{\bar{\theta}} \rangle
& = & \frac{J_2 (K) }{2},
\earr
where, given $\theta= \theta(t)$, $\bar{\theta}=\theta(t+1)$
and $\bar{\bar{\theta}}=\theta(t+2)$,
$J_2 (K)$ is the Bessel function of the first kind of index 2
and $K$ is the classical chaos parameter.
Correlations between more distant kicks, 
$\langle \cos [\theta(t)] \cos [\theta(t^\prime)] \rangle$, 
with $t^\prime - t >2$, 
though non vanishing, are very
weak for $K\gg 1$, therefore hereafter we will neglect them.
Correlations in the kicked rotator dynamics 
eventually result in a modification of the decay
rate $\Gamma$ of entanglement, as it is clearly shown in Fig.~\ref{fig:krate},
where we plot the dependence of $\Gamma$ as a function of $K$.

In order to study analytically memory effects on the entanglement decay, we 
provide a simple generalization of the random phase-kick model 
(\ref{eq:randomphase}), so that the correlations 
of Eqs.~(\ref{eq:corr1})-(\ref{eq:corr2}) are taken into account.
We use the following conditional probability distribution
for the angle $\bar{\theta}$ at time $t + 1$, given the 
angle $\theta$ at time $t$:
\beq
p(\bar{\theta} \vert \theta) =
\left\{ \begin{array}{ll} \displaystyle
\sqrt{J_2} \: \delta \hspace{-1mm} \left[ \bar{\theta} - \theta +
\frac{\pi}{2} (-1)^t \right] + \frac{1 - \sqrt{J_2}}{2 \pi}, &
(J_2 > 0), \vspace{1mm} \\ \displaystyle
\sqrt{-J_2} \: \delta \hspace{-1mm} \left[ \bar{\theta} - \theta -
\frac{\pi}{2} \right] + \frac{1 - \sqrt{-J_2}}{2 \pi}, & (J_2 < 0).
\end{array} \right. \label{eq:probcorr}
\eeq
This distribution corresponds to having a $\bar{\theta}$ angle which is
correlated with $\theta$ (i.e. $\bar{\theta} = \theta \pm \pi / 2$)
with probability $p_{\mathrm{c}} = \sqrt{\abs{J_2}}$, and completely
uncorrelated with probability $p_{\mathrm{nc}} = 1-p_{\mathrm{c}}$.
For our purposes, the relevant properties of the probability 
distribution $p(\theta,\bar{\theta})=p(\theta)p(\bar{\theta}|\theta)$ are:
$\int d \theta \int d \bar{\theta} \, p(\theta, \bar{\theta}) = 1$;
$\int d \theta \int d \bar{\theta} \, p(\theta, \bar{\theta}) \cos \theta
\cos \bar{\theta} = 0$;
$\int d \theta \int d \bar{\theta} \, p(\theta, \bar{\bar{\theta}})
\cos \theta \cos \bar{\bar{\theta}} = J_2/2$.
Therefore, Eq.~(\ref{eq:probcorr}) is a convenient 
distribution probability leading to a decay of the 
bath correlation function as in
Eqs.~(\ref{eq:corr1})-(\ref{eq:corr2}) for the kicked rotator model.
Of course, other probability distributions could equally well reproduce 
such decay.

Given the probability distribution $p(\bar{\theta}|\theta)$, 
we can replace \eqref{eq:randomphase} with the following
two-kicks time evolution map:
\beq \begin{array}{l}
\bar{\bar{\rho}}_{12} = \displaystyle 
\int_{0}^{2 \pi} \hspace{-1.5mm} d\theta \,
\int_{0}^{2 \pi} \hspace{-1.5mm} d \bar{\theta} \, p(\theta,\bar{\theta}) \,
\hat{R} (\bar{\theta})  \hat{R} ({\theta}) \, \rho_{12} \,
\hat{R}^{\dagger} ({\theta}) \hat{R}^{\dagger} (\bar{\theta}),
\end{array}
\label{eq:corrphase}
\eeq
where $\rho_{12}$ and $\bar{\bar{\rho}}_{12}$ denote the two-qubit density matrix 
at times $t$ and $t+2$ and 
$p(\theta,\bar{\theta})$ is the joint probability to have a rotation
through an angle $\theta$ at a time $t$ and through an angle $\bar{\theta}$ 
at time $t+1$.
Therefore, $p(\theta,\bar{\theta})$ 
accounts for correlations between the angles at subsequent kicks.
Clearly, if angles are completely uncorrelated we have
$p(\theta,\bar{\theta})=1/4 \pi^2$,
thus recovering the random phase-kick model \eqref{eq:randomphase}.
Note that, for the sake of simplicity, map (\ref{eq:corrphase}) has been written
for $\delta_1=\delta_2=0$ (the generalization of this map to 
$\delta_1 , \delta_2 \ne 0$ is straightforward). 

The phase-kick model (\ref{eq:corrphase}) can be simulated 
by using the quantum trajectories approach \cite{carmichael,brun,carlo}. 
This method is very convenient
in the study of dissipative systems: instead of solving a master equation,
one stochastically evolves a state vector, and then averages over many runs.
At the end, we get the same probabilities as the ones directly obtained 
through the density matrix.
In our case, the effect of the kicked rotator on the two-qubits
wave function is simply that of a rotation through an angle whose value
is drawn according to the probability distribution (\ref{eq:probcorr}).
Numerical data obtained with the quantum trajectories method are plotted 
in Fig.~\ref{fig:krate} (stars); notice that, at $K\gg 1$, they are in good agreement 
with data from simulation of the Hamiltonian model (\ref{eq:hammodel}) 
(circles). 

It is also possible to give an analytic estimate of the decay rate 
$\Gamma(K)$ in the limit in which the free evolution of the two qubits
can be neglected (i.e., we take $\delta_1 ,\delta_2 \ll \epsilon \ll 1$).
As for the random phase model \eqref{eq:randomphase}, in this limit
the effect of map~\eqref{eq:corrphase} is pure dephasing.
At every map step the density matrix $\rho_{12}$ is of the form of
Eq.~\eqref{eq:rhodeco}. The coherences $D_t$ can be computed 
by iterating Eq.~\eqref{eq:corrphase}. We obtain what follows for the
first time steps:
\beq
\left\{ \begin{array}{l}
D_1 = 1 - 4 \epsilon^2, \vspace{0.3mm} \\
D_2 = 1 - 8 \epsilon^2, \vspace{0.3mm} \\
D_3 = 1 - \big[ 12 + 8 J_2 (K) \big] \epsilon^2 ,\vspace{0.3mm} \\
D_4 = 1 - \big[ 16 + 16 J_2 (K) \big] \epsilon^2. \vspace{0.3mm}
\end{array} \right. \label{eq:corranalyt}
\eeq
Assuming an exponential decay of entanglement, $E_{12}(t)\sim e^{-\Gamma t}$,
we can evaluate $\Gamma$ starting from $D_3$ and $D_4$:
$\Gamma \approx \ln ( E_3 / E_4 )$.
Thus, from Eqs.~\eqref{eq:corranalyt} we obtain the following
analytic estimate:
\beq
\Gamma \approx \frac{4 \epsilon^2}{\ln 2} \left( 1 + 2 J_2 (K) \right).
\label{eq:corrrate} \eeq

In the inset of Fig.~\ref{fig:krate} we compute the entanglement decay
rates as a function of $K$ in the case $\delta_1 = \delta_2 = 0$.
This figure clearly shows that the 
rates obtained from numerical data for the chaotic environment model
(circles) are in good agreement, when $K\gg 1$, with the analytic estimate
\eqref{eq:corrrate} (solid curve).

We point out that the oscillations with $K$ of the entanglement decay rate 
$\Gamma(K)$ are ruled by the Bessel function $J_2(K)$, in the same way as for
the well-known $K$-oscillations of the classical diffusion coefficient $D(K)$
\cite{rechester80}. Therefore, the entanglement decay rate is strictly related to
a purely classical quantity. The ultimate reason of this relation is rooted 
in the fact that the bath correlation function relevant for the study of memory 
effects,  
$\langle \cos [\theta(t)] \cos [\theta(t^\prime)] \rangle$,
also determines the deviations of the diffusion coefficient $D(K)$ from the random 
phase approximation value $K^2/2$.

\section{Conclusions} \label{sec:conclusions}

We have shown that a single particle in the chaotic regime 
can be equivalent to a bath with infinitely many degrees of freedom 
as a dephasing environment. This provides a clear example of the 
relevance of chaos to decoherence. Furthermore, also memory
effects can be included in a single particle chaotic environment.
This observation paves the way to a convenient numerical simulation
of non-Markovian dynamics. Indeed, both exponential and power-law
decays of the bath (single particle) correlation functions 
can be obtained in chaotic maps \cite{sawtooth,liverani}.
Therefore, these maps could be used to efficiently simulate 
important quantum noise models such as 
random telegraphic or $1/f$ noise.    

\begin{acknowledgments}
We thank R. Artuso for very useful discussions.
This work was supported in part by the PRIN 2005 
``Quantum computation with trapped particle arrays, neutral and 
charged'', the MIUR-PRIN, and the European Community
through grants RTNNANO and EUROSQIP. The present work has been
performed within the Quantum Information research program
of Centro di Ricerca Matematica ``Ennio De Giorgi''
of Scuola Normale Superiore.
\end{acknowledgments}

\appendix

\section{The random phase model} \label{app1}

Here we provide an explicit set of equations describing 
map \eqref{eq:randomphase} in the Bloch representation.
In such representation, a generic two-qubit mixed state can
be written as
\beq
\rho_{12} = \frac{1}{4} \hat{\mathbb{I}} \otimes \hat{\mathbb{I}} +
\sum_{i=1}^3 \alpha_i ( \hat{\sigma}_i \otimes \hat{\mathbb{I}} )+
\sum_{j=1}^3 \beta_j ( \hat{\mathbb{I}} \otimes \hat{\sigma}_j ) +
\sum_{i,j=1}^3 \gamma_{ij} (\hat{\sigma}_i \otimes \hat{\sigma}_j)
\eeq
We insert this expansion into Eq.~\eqref{eq:randomphase}
and evaluate the commutators between $\rho_{12}$ and the terms that multiply 
$\rho_{12}$ on the right in \eqref{eq:randomphase} 
($e^{i \delta_1 \hat{\sigma}_x^{(1)}}$, $e^{i \delta_2 \hat{\sigma}_x^{(2)}}$,
and $\hat{R}^{\dagger} (\theta)$).
For this purpose, we use the standard commutation rules
for the Pauli matrices:
\beq
\left[ \hat{\sigma}_i, \hat{\sigma}_j \right] =
2 i \, \epsilon_{ijk} \, \hat{\sigma}_k \, ,
\eeq
where $\epsilon_{ijk}$ is the Levi-Civita tensor, 
which is equal to $+1$ if $ijk$ is an even permutation
of $123$, to $-1$ for odd permutations and $0$ otherwise.
We also use the identity
\beq
e^{i \delta \hat{\sigma}_x} = (\cos \delta) \, \hat{\mathbb{I}} + 
i \, (\sin \delta) \, \hat{\sigma}_x
\eeq
and expand the exponentials in $\hat{R}^\dagger (\theta)$ to the second
order in $\epsilon$.
The average over $\theta$ is then performed using
\beq \begin{array}{c}
\left< \cos^{2n+1} \theta \right>_\theta = 0, \vspace{2mm} \\ \qquad
\left< \cos^{2} \theta \right>_\theta = \frac{1}{2}, \qquad
\left< \cos^{4} \theta \right>_\theta = \frac{3}{8}.
\end{array}
\eeq

In this way we arrive at the following one-kick map, valid
in the approximation 
$\epsilon, \delta_1, \delta_2  \ll 1$:

\beq \begin{array}{c}

\left\{ \begin{array}{l}
\bar{\alpha}_x = \alpha_x (1 - \epsilon^2), \\
\bar{\alpha}_y = \alpha_y (1 - \epsilon^2) - 2 \delta_1 \alpha_z,\\
\bar{\alpha}_z = \alpha_z + 2 \delta_1 \alpha_y,
\end{array} \right.

\vspace{2mm} \\

\left\{
\begin{array}{l}
\bar{\beta}_x = \beta_x (1 - \epsilon^2), \\
\bar{\beta}_y = \beta_y (1 - \epsilon^2) - 2 \delta_2 \beta_z,\\
\bar{\beta}_z = \beta_z + 2 \delta_2 \beta_y,
\end{array} \right.

\vspace{2mm} \\

\left\{ \begin{array}{l}
\bar{\gamma}_{xx} = \gamma_{xx} (1 - 2 \epsilon^2) +
2\epsilon^2 \gamma_{yy}, \\
\bar{\gamma}_{yy} = \gamma_{yy} (1 - 2 \epsilon^2) +
2\epsilon^2 \gamma_{xx} - 2 \delta_1 \gamma_{zy}
- 2 \delta_2 \gamma_{yz}, \\
\bar{\gamma}_{zz} = \gamma_{zz} + 2 \delta_1 \gamma_{yz}
+ 2 \delta_2 \gamma_{zy}, \\
\bar{\gamma}_{xy} = \gamma_{xy} (1 - 2 \epsilon^2) -
2 \epsilon^2 \gamma_{yx} - 2 \delta_2 \gamma_{xz}, \\
\bar{\gamma}_{yx} = \gamma_{yx} (1 - 2 \epsilon^2) -
2 \epsilon^2 \gamma_{xy} - 2 \delta_2 \gamma_{zx}, \\
\bar{\gamma}_{xz} = \gamma_{xz} (1 - 2 \epsilon^2) +
2 \delta_2 \gamma_{xy}, \\
\bar{\gamma}_{zx} = \gamma_{zx} (1 - 2 \epsilon^2) +
2 \delta_1 \gamma_{yx}, \\
\bar{\gamma}_{yz} = \gamma_{yz} (1 - 2 \epsilon^2) -
2 \delta_1 \gamma_{zz} + 2 \delta_2 \gamma_{yy}, \\
\bar{\gamma}_{zy} = \gamma_{zy} (1 - 2 \epsilon^2) +
2 \delta_1 \gamma_{yy} - 2 \delta_2 \gamma_{zz}.
\end{array} \right.

\label{eq:mapequations} \end{array}
\eeq
Note that these equations can be straightforwardly
generalized to the continuous time limit.

We now focus on the case in which the two qubits are initially
in the Bell state $\ket{\phi^+}$, so that the only non zero coordinates
in the Bloch representation of $\rho_{12}(t=0) = \ket{\phi^+} \bra{\phi^+}$
are $\gamma_{xx} = \gamma_{zz} = - \gamma_{yy} = 1/4$.
The system in Eq.~\eqref{eq:mapequations} then reduces to a set of
five coupled equations for the $\gamma$'s:
\beq
\left\{ \begin{array}{l}
\bar{\gamma}_{xx} = \gamma_{xx} (1 - 2 \epsilon^2) +
2\epsilon^2 \gamma_{yy}, \\
\bar{\gamma}_{yy} = \gamma_{yy} (1 - 2 \epsilon^2) +
2\epsilon^2 \gamma_{xx} - 2 \delta_1 \gamma_{zy}
- 2 \delta_2 \gamma_{yz}, \\
\bar{\gamma}_{zz} = \gamma_{zz} + 2 \delta_1 \gamma_{yz}
+ 2 \delta_2 \gamma_{zy}, \\
\bar{\gamma}_{yz} = \gamma_{yz} (1 - 2 \epsilon^2) -
2 \delta_1 \gamma_{zz} + 2 \delta_2 \gamma_{yy}, \\
\bar{\gamma}_{zy} = \gamma_{zy} (1 - 2 \epsilon^2) +
2 \delta_1 \gamma_{yy} - 2 \delta_2 \gamma_{zz}.
\end{array} \right. \label{eq:mapreduced}
\eeq
Their solution after $t$ map steps is given,
in the limiting case $\delta_1 , \delta_2 \ll \epsilon \ll 1$, by 
\beq
\gamma_{xx}(t) \approx - \gamma_{yy}(t) \approx \frac{1}{4} - \epsilon^2 t, \quad
\gamma_{zz}(t) \approx \frac{1}{4}.
\eeq
Note that here we also require $\epsilon^2 t \ll 1$.
The corresponding density matrix is
\beq
\rho_{12}(t) = \frac{1}{2} \left( \begin{array}{cccc}
1 & 0 & 0 & D_t \\ 0 & 0 & 0 & 0 \\ 0 & 0 & 0 & 0 \\ D_t & 0 & 0 & 1
\end{array} \right), \label{eq:rhodeco}
\eeq
with 
\beq
D_t = 1 - 4 \epsilon^2 t.
\eeq
The calculation of the von Neumann entropy $S_{12}(t)$ and of the entanglement $E_{12}(t)$
for state~\eqref{eq:rhodeco} is straightforward and leads to 
Eqs.~\eqref{eq:Eestim}-\eqref{eq:Sestim}.

\section{Angular correlations in the Chirikov standard map} \label{app2}

After rescaling the angular momentum $n \to p = Tn$, the 
Chirikov standard map \eqref{eq:chirikov} becomes
\beq
\left\{ \begin{array}{l}
\bar{p} = p + K \sin \theta, \\
\bar{\theta} = \theta + \bar{p}, 
\hspace{1.cm} (\textrm{mod } 2 \pi),
\end{array} \right. \label{eq:chirikov2}
\eeq
where $K=kT$. 
In this appendix we provide a simple proof of
Eqs.~\eqref{eq:corr1}-\eqref{eq:corr2}, namely we evaluate correlations between
the cosines of the phases of two consecutive and next-consecutive kicks.
Averages are performed over all the input phase space:
\beq
\left< f(\theta,p) \right> \equiv \lim_{P \to \infty} \left\{ \int_{-P}^P
\frac{\rm{d}p}{2P} \: \int_0^{2 \pi} \frac{f(\theta,p)}{2\pi} \, \rm{d}\theta \right\}
\eeq
By using Eq.~\eqref{eq:chirikov2}, we obtain the following equalities for correlations
between two consecutive kicks:
\beq \begin{array}{l}
\displaystyle \hspace*{-0.5cm} \left< \cos \theta 
\cos \bar{\theta} \right> =
\left< \cos \theta 
\cos \left( \theta + p + K \sin \theta \right) \right>
\vspace{1mm} \\
\displaystyle \hspace*{0.5cm} = \left< \cos p \right>
\left< \cos \theta 
\cos \left( \theta + K \sin \theta \right)
\right> \vspace{1mm} \\ \displaystyle \hspace*{0.5cm}
- \, \left< \sin p \right> 
\left< \cos \theta 
\sin \left( \theta + K \sin \theta \right) \right> =0 ,
\end{array} \eeq
where we have used some trivial trigonometric identities, and $\left< \cos x \right> =
\left< \sin x \right> = 0$.

Correlations between two next-consecutive kicks are evaluated by considering
values at times $t+1$ (denoted by an overbar) and $t-1$ (denoted by an underbar).
The map in Eq.~\eqref{eq:chirikov2} can be straightforwardly inverted
to give backward evolution:
\beq
\left\{ \begin{array}{l}
\underline{p}= p -  K \sin (\theta-p), \\
\underline{\theta}= \theta - p, \hspace{1.cm} (\textrm{mod } 2 \pi).
\end{array} \right. \label{eq:chirikov3}
\eeq
Therefore, from Eqs.~\eqref{eq:chirikov2} and \eqref{eq:chirikov3}, after some
simple algebra we obtain:
\beq \begin{array}{l}
\displaystyle \left< \cos \bar{\theta} 
\cos \underline{\theta} \right> =
\left< \cos \left( \theta + p + K \sin \theta \right) 
\cos \left( \theta - p \right) \right> \\
\displaystyle = \frac{1}{2} \left< \cos \left( 2 \theta + K \sin \theta
\right) \right> = \frac{1}{2} J_{-2} (K) = \frac{1}{2} J_2 (K),
\end{array}
\eeq
where
\beq
J_n (K) = \frac{1}{2 \pi} \int_0^{2 \pi}
\cos \left( K \sin \theta - n \theta \right) \textrm{d} \theta
\eeq
is the Bessel function of the first kind of index 2.



\begin{thebibliography}{99}
\bibitem{note_frequencies}
  In the limit in which one of the two qubits' oscillation frequencies
  is much smaller than the other, the period $T_{\mathrm{osc}}$
  of entanglement oscillations is given by
  $T_{\mathrm{osc}} \sim \pi / \mathrm{max}(\delta_1, \delta_2)$.
  When the two frequencies are comparable ($\delta_1 \sim \delta_2$),
  then $T_{\mathrm{osc}} \sim \pi / \bar{\delta}$, where
  $\bar{\delta} = (\delta_1 + \delta_2 )/2$.
  In general $T_{\mathrm{osc}}$ is a non linear function
  of $\delta_1$ and $\delta_2$; this feature is contained in
  Eqs.~\eqref{eq:mapequations}, which have to be solved numerically
  for each specific case.
\end{thebibliography}
\end{document}